\documentclass[twocolumn,showpacs,preprintnumbers,amsmath,amssymb]{revtex4}

\usepackage{graphicx}
\usepackage{dcolumn}
\usepackage{bm}

\begin{document}

\preprint{APS/213-QED}

\title{Enhancement of Field Squeezing Using Coherent Feedback}

\author{J.E. Gough}
 \email{jug@aber.ac.uk}
 
\author{S. Wildfeuer}
 \email{sew08@aber.ac.uk}
 
\affiliation{Institute for Mathematics and Physics, Aberystwyth University, SY23 3BZ, Wales, United Kingdom.}

\date{\today}

\begin{abstract}

The theory of quantum feedback networks has recently been developed with the aim of showing how quantum input-output components may be connected together so as to control, stabilize or enhance the performance of one of the subcomponents. In this paper we show how the degree to which an idealized component (a degenerate parametric amplifier in the strong-coupling regime) can squeeze input fields may be enhanced by placing the component in-loop in a simple feedback mechanism involving a beam splitter. We study the spectral properties of output fields, placing particular emphasis on the elastic and inelastic components of the power density.
\end{abstract}

\maketitle

\section{Introduction}

In the last two decades quantum physics has witnessed a remarkable convergence between theoretical models of interactions, particularly for open systems and measurement apparatuses, and experimental implementations of quantum engineering. The unifying framework has been to import and adapt the principles of control theory to the quantum domain.
The advantages over traditional approaches show that quantum control will play a fundamental role in emerging quantum technologies \cite{MK05,DM03}. A variety of promising control techniques have been put forward \cite{PDR88} -\cite{K99} which extend open-loop paradigms (where control inputs are decided in advance) and measurement-based closed-loop paradigms (where feedback of observations is used to determine the control inputs). Real-time measurement-based feedback has been applied to adaptive homodyne measurement \cite{W95, AASDM02} to achieve measurement variances close to the standard quantum limit. 

Our interest lies in coherent quantum control, which is a non-measurement based feedback approach. Quantum feedback networks \cite{GJ09,GJ08b} have emerged as a natural class of objects with which to address assemblies of quantum input-output components so as to allow feedforward and feedback connections. This offers a convenient framework to formulate problems in coherent quantum control and robust quantum control problems \cite{M08}-\cite{JNP08}. (We remark that the early formulation of coherent quantum feedback control due to Lloyd \cite{NWCL00} deals with the direct interaction between system and its controller, as opposed to one mediated by quantum field processes. However, this may be treated as a special case of the network \cite{GJ08b}.)

An early application of feedback to enhance the squeezing of an (infrared) cavity mode was given by Wiseman \emph{et al}. \cite{WTB95}. Here the mode is coupled to a second harmonic (green) mode which is subjected to a quantum
nondemolition measurement. In contrast, we wish to examine the squeezing of the input noise field by a cavity mode acting as an idealized squeezing device. Here the feedback is coherent, rather than measurement-based, and we consider a set up involving a simple beam splitter to introduce the feedback loop. We shall work in the limit of instantaneous feedback throughout. We shall be interested in the class of linear dynamical systems \cite{YK03a},\cite{GGY08},\cite{GJN09},
and indeed will study static components wherein the internal degrees of freedom have been eliminated.

\begin{figure}
	\centering
		\includegraphics[width=0.65\linewidth]{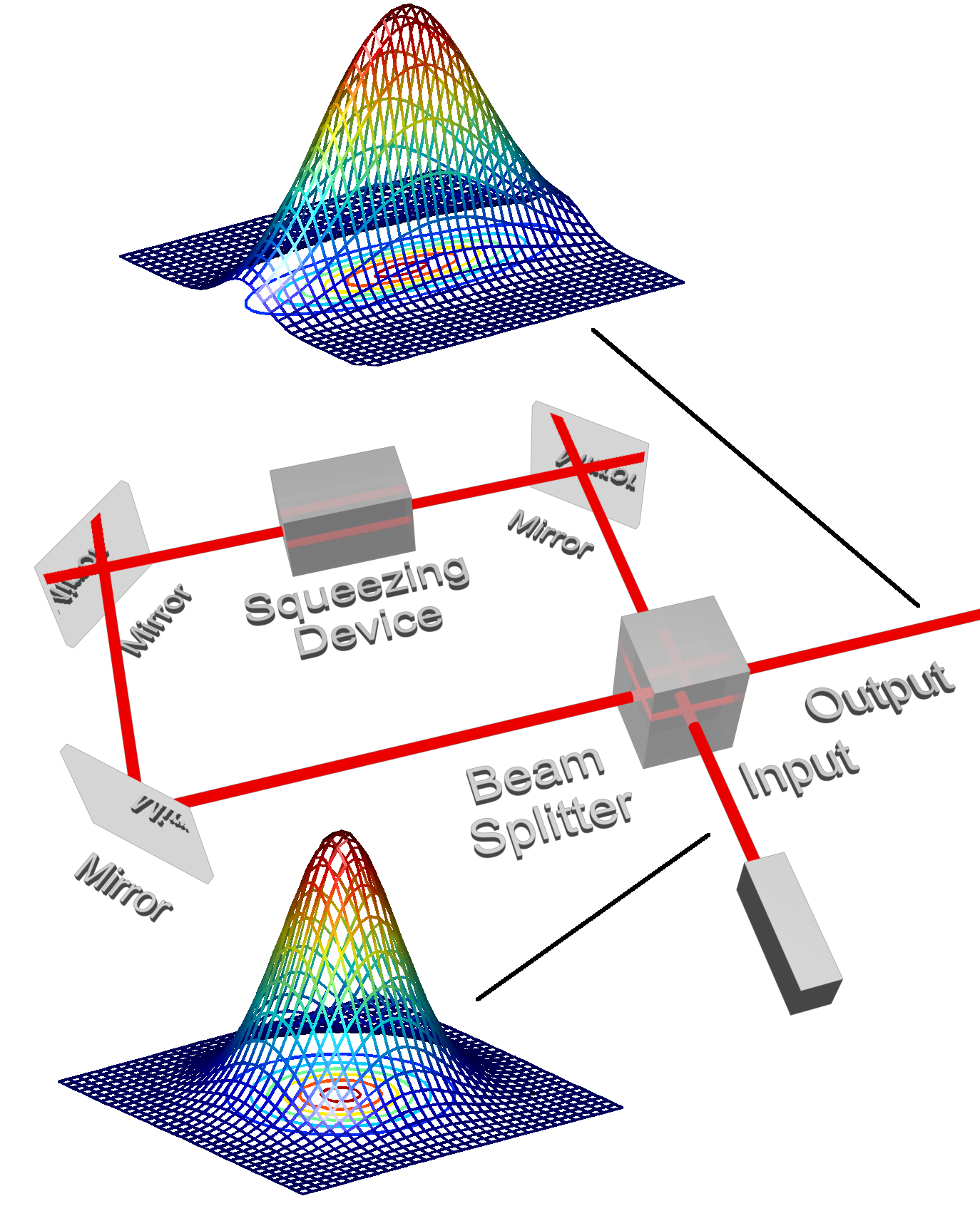}
	\label{picture}
	\caption{(Color online) Squeezing Device in a Feedback Loop}
\end{figure}

The degenerate parametric amplifier (DPA) is a well known non-linear device capable of squeezing input fields \cite{LYS61} -\cite{NB95}. We follow the treatment of Gardiner \cite{GZ00}. For a single quantum input field coupled to a single cavity mode $a$ with coupling strength $\sqrt{\kappa }$ and Hamiltonian 
\begin{equation}
H_{\text{DPA}}=\frac{i\varepsilon }{4}\left( a^{\ast 2}-a^{2}\right) ,
\label{H_DPA}
\end{equation}
there is an approximate squeezing parameter given by, \cite{GZ00} section
7.2.9, 
\begin{equation}
r_{\text{DPA}}=\ln \left( \frac{\kappa +\varepsilon }{\kappa -\varepsilon }%
\right) .  \label{r_open_loop}
\end{equation}
Here the amplification is due to the specific choice of the Hamiltonian $H_{\text{DPA}}$.  

Without feedback, the method of obtaining maximal squeezing for a degenerate parametric amplifier is to try and realize the Hamiltonian for the internal mode with parameter coefficient $\varepsilon $ as close to the threshold
value $\left( \varepsilon =\kappa \right) $ as possible, see \cite{GZ00} section 10.2. As originally noted by Yanagisawa and Kimura \cite{YK03a}, the value of the effective damping for an in-loop mode, see FIG. \ref{picture} will depend on the reflectivity value $\alpha  $:
\begin{equation}
\kappa \left( \alpha \right) =\frac{1-\alpha }{1+\alpha }\kappa .
\label{kappa_eff}
\end{equation}
Our strategy is to use coherent feedback for a fixed degenerate parametric amplifier (below threshold, and therefore internally stable \cite{GJN09})and tune the reflectivity of the beam splitter so as to select the degree of squeezing.

The degenerate parametric amplifier is an idealized device in which one assumes that $\kappa $ and $\varepsilon $ are large but with fixed ratio. We shall investigate the situation where both these parameters are finite. Also, we introduce additional quantum damping into the model to see the effect of loss.

\section{Quantum Feedback Networks}

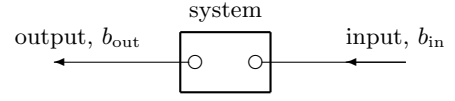
\begin{figure}
\begin{center}
\setlength{\unitlength}{.04cm} 
\begin{picture}(120,45)
\label{pic1}
\thicklines
\put(45,10){\line(0,1){20}}
\put(45,10){\line(1,0){30}}
\put(75,10){\line(0,1){20}}
\put(45,30){\line(1,0){30}}
\thinlines
\put(48,20){\vector(-1,0){45}}
\put(120,20){\vector(-1,0){20}}
\put(120,20){\line(-1,0){48}}
\put(50,20){\circle{4}}
\put(70,20){\circle{4}}
\put(100,26){input, $b_{\rm in}$}
\put(48,35){system}
\put(-10,26){output, $ b_{\rm out}$}
\end{picture}
\caption{Input-output component}
\label{i_o_component}
\end{center}
\end{figure}

A single component consists of a quantum mechanical system, with Hilbert space $\mathfrak{h}$ driven by $n$ quantum input processes $b_{\text{in},i} \left( i=1,\cdots ,n\right) $, \cite{GC85, GZ00}, satisfying canonical
commutation relations of the form $[b_{\text{in},i}(t),b_{\text{in} ,j}(t^{\prime })]=0$, $[b_{\text{in},i}^{\ast }(t),b_{\text{in},j}^{\ast }(t^{\prime })]=0$ and 
\begin{equation}
\lbrack b_{\text{in},i}\left( t\right) ,b_{\text{in},j}^{\ast }\left(
t^{\prime }\right) ]=\delta _{ij}\,\delta \left( t-t^{\prime }\right) .
\label{CCR}
\end{equation}
A schematic of a component appears in FIG. \ref{i_o_component}.

The component is characterized by \textit{generator} $G=\left( S,L,H\right) $ where $S=\left( S_{ij}\right) $ is a unitary $n\times n$ matrix whose entries are operators on $\mathfrak{h}$ called the \textit{scattering
coefficient matrix}, $L=\left( L_{i}\right) $ is a column vector whose entries are operators on $\mathfrak{h}$ called the \textit{coupling coefficient vector}, and $H$ is a self-adjoint operator on $\mathfrak{h}$ giving the system
\textit{\ Hamiltonian}. On the joint system-field space we have the unitary evolution process $U\left( t\right) $ which satisfies the quantum It\={o} QSDE \cite{HP84}

\begin{multline}
dU\left( t\right) =\left\{ \sum_{i,j}\left( S_{ij}-\delta _{ij}\right)
d\Lambda _{\text{in},ij}\left( t\right) +\sum_{i}L_{i}dB_{\text{in},i}^{\ast
}\left( t\right) \right.   
\\
\left. -\sum_{i,j}L_{i}^{\ast }S_{ij}dB_{\text{in},j}\left( t\right) -(\frac{%
1}{2}\sum_{i}L_{i}^{\ast }L_{i}+iH)dt\right\} U\left( t\right) ,
\label{U}
\end{multline}
with $U(0)=I$.

We encounter the integrated fields $B_{\text{in},i}\left( t\right)
=\int_{0}^{t}b_{\text{in},i}\left( t^{\prime }\right) dt^{\prime }$, $B_{%
\text{in},j}^{\ast }\left( t\right) =B_{\text{in},j}\left( t\right) ^{\ast }$%
\ and $\Lambda _{\text{in},ij}\left( t\right) =\int_{0}^{t}b_{\text{in}%
,i}^{\ast }\left( t^{\prime }\right) b_{\text{in},j}\left( t^{\prime
}\right) dt^{\prime }$ which satisfy the following quantum It\={o} table 
\cite{HP84} 
\begin{equation*}
\begin{tabular}{l|llll}
$\times $ & $dB_{\text{in},j}$ & $d\Lambda _{\text{in},jl}$ & $dB_{\text{in}%
,j}^{\ast }$ & $dt$ \\ \hline
$dB_{\text{in},i}$ & 0 & $\delta _{ij}dB_{\text{in},l}$ & $\delta _{ij}dt$ & 
0 \\ 
$d\Lambda _{\text{in},ki}$ & 0 & $\delta _{ij}d\Lambda _{\text{in},kl}$ & $%
\delta _{ij}dB_{\text{in,}k}^{\ast }$ & 0 \\ 
$dB_{\text{in},i}^{\ast }$ & 0 & 0 & 0 & 0 \\ 
$dt$ & 0 & 0 & 0 & 0
\end{tabular}
.
\end{equation*}

\subsection{Components In Loop}

We may consider a feedback arrangement using a beam splitter, as in FIG. \ref{FBS} below. The beam splitter is a static device which we take to be described by

\begin{equation*}
\left( 
\begin{array}{c}
b_{\text{out}} \\ 
v_{\text{in}}
\end{array}
\right) =T\left( 
\begin{array}{c}
b_{\text{in}} \\ 
v_{\text{out}}
\end{array}
\right) ,\qquad T=\left( 
\begin{array}{cc}
\alpha & \beta \\ 
\mu & \nu
\end{array}
\right) ,
\end{equation*}
where $T$ is taken to be a real-valued unitary matrix with determinant $%
\sigma =\alpha \nu -\beta \mu =\pm 1$.

Suppose that $G_{1}^{0}=(S_{1}^{0},L_{1}^{0},H_{1}^{0})$ is the generator of
the $\left( n=1\right) $ component before the feedback connections are made.
Once the component is in loop, in the limit of instantaneous feedback, we
find an effective component with input $b_{\text{in}}$ and $b_{\text{out}}$
as indicated and generator given by $G_{1}=\left( S_{1},L_{1},H_{1}\right) $
with \cite{GJ09} 
\begin{eqnarray}
S_{1} &=&\alpha +\beta \left( \left( S_{1}^{0}\right) ^{-1}-\nu \right) \mu ,
\\
L_{1} &=&\beta \left( 1-\nu S_{1}^{0}\right) ^{-1}L_{1}^{0}, \\
H_{1} &=&H_{1}^{0}+\text{Im}\left\{ \left( L_{1}^{0}\right) ^{\ast }\left(
1-\nu S_{1}^{0}\right) ^{-1}L_{1}^{0}\right\} .
\end{eqnarray}

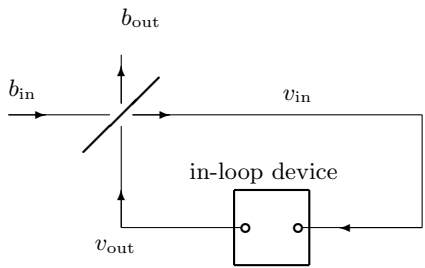
\begin{figure}
\begin{center}
\setlength{\unitlength}{.05cm} 
\begin{picture}(120,80)

\thicklines
\put(30,40){\line(1,1){20}}

\thinlines
\put(10,50){\line(1,0){27}}
\put(10,50){\vector(1,0){10}}
\put(40,20){\line(0,1){27}}
\put(40,20){\vector(0,1){10}}
\put(40,53){\line(0,1){13}}
\put(40,53){\vector(0,1){10}}
\put(43,50){\line(1,0){77}}
\put(43,50){\vector(1,0){10}}
\put(120,20){\line(0,1){30}}
\put(40,20){\line(1,0){32}}
\put(120,20){\line(-1,0){32}}
\put(120,20){\vector(-1,0){22}}

\thicklines

\put(70,10){\line(0,1){20}}
\put(70,10){\line(1,0){20}}
\put(90,10){\line(0,1){20}}
\put(70,30){\line(1,0){20}}
\put(87,20){\circle{2}}
\put(73,20){\circle{2}}
\put(56,33){ in-loop device}

\put(10,55){${ b}_{\rm in}$}
\put(40,74){${ b}_{\rm out}$}
\put(83,54){${ v}_{\rm in}$}

\put(33,14){${ v}_{\rm out}$}
\end{picture}
\caption{Feedback using a beam-splitter}
\label{FBS}
\end{center}
\end{figure}

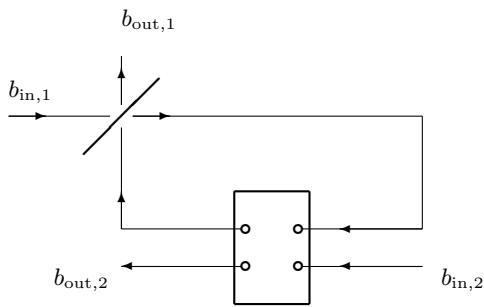
\begin{figure}
\begin{center}
\setlength{\unitlength}{.05cm} 
\begin{picture}(130,90)

\thicklines
\put(30,50){\line(1,1){20}}

\thinlines
\put(10,60){\line(1,0){27}}
\put(10,60){\vector(1,0){10}}
\put(40,30){\line(0,1){27}}
\put(40,30){\vector(0,1){10}}
\put(40,63){\line(0,1){13}}
\put(40,63){\vector(0,1){10}}
\put(43,60){\line(1,0){77}}
\put(43,60){\vector(1,0){10}}
\put(120,30){\line(0,1){30}}
\put(40,30){\line(1,0){32}}
\put(120,30){\line(-1,0){32}}
\put(120,30){\vector(-1,0){22}}

\put(40,20){\line(1,0){32}}
\put(50,20){\vector(-1,0){10}}
\put(120,20){\line(-1,0){32}}
\put(120,20){\vector(-1,0){22}}

\thicklines

\put(70,10){\line(0,1){30}}
\put(70,10){\line(1,0){20}}
\put(90,10){\line(0,1){30}}
\put(70,40){\line(1,0){20}}
\put(87,20){\circle{2}}
\put(73,20){\circle{2}}
\put(87,30){\circle{2}}
\put(73,30){\circle{2}}

\put(10,65){${ b}_{{\rm in},1}$}
\put(40,84){${ b}_{{\rm out},1}$}

\put(22,15){${b}_{{\rm out},2}$}
\put(125,15){${b}_{{\rm in},2}$}
\end{picture}
\caption{Feedback with loss}
\label{FWL}
\end{center}
\end{figure}

The example above may be extended to include a loss mechanism describing
coupling of the component to the environment, see FIG. \ref{FWL}. Prior to making
connections we assume that the component is the four-port system $\left(
n=2\right) $ with generator given by 
\begin{equation*}
S_{0}=\left( 
\begin{array}{cc}
S_{1}^{0} & 0 \\ 
0 & 1
\end{array}
\right) ,L_{0}=\left( 
\begin{array}{c}
L_{1}^{0} \\ 
L_{2}^{0}
\end{array}
\right) .
\end{equation*}
After feedback, the effective generator becomes $G=\left( S,L,H\right) $
with 
\begin{equation*}
S=\left( 
\begin{array}{cc}
S_{1} & 0 \\ 
0 & 1
\end{array}
\right) ,L=\left( 
\begin{array}{c}
L_{1} \\ 
L_{2}^{0}
\end{array}
\right) ,H=H_{1}.
\end{equation*}
In the language of \cite{GJ09},\cite{GJ08b} the effective generator is the
concatenation $G=\left( S_{1},L_{1},H_{1}\right) \boxplus \left(
1,L_{2}^{0},0\right) $. We have reasoned that since $S_{0}$ is diagonal,
there is no direct scattering between the inputs to the in loop device, and
that concatenation of the effective lossless generator $G_{1}$ with the loss
mechanism $G_{2}^{0}$. This however can be shown to be correct by utilizing
the following construction from \cite{GJ09}: we note that the beam splitter
itself can be understood as a static four-port component $G^{\prime }=\left(
T,0,0\right) $ and the set up in FIG. \ref{FWL} is then naturally identified as a
Redheffer star-product arrangement of the two four-port devices, the
effective generator for components in a Redheffer formation is given in
section 5.3 of \cite{GJ09}, and substitution into the expression gives
precisely the generator $G$.

We note that the relations we shall derive below for linear systems can be
arrived at by algebraically eliminating the internal fields $v_{\text{in}}$
and $v_{\text{out}}$. Whilst this is obviously easier than evoking the
mathematical formulation of quantum feedback networks, we should point out
that this is not entirely consistent and that the in-loop fields $v_{\text{in%
}}$ and $v_{\text{out}}$ are not canonical! Whilst this has been incorrectly
interpreted elsewhere as a violation of the Heisenberg uncertainty
relations, the reality is that the above description emerges from a regular
model in which the commutation relations hold at all times however the
finite time delay of the feedback is taken into account \cite{GJ09}. The
in-loop fields are eliminated in the instantaneous feedback limit and should
then not be thought of as real physical fields. The algebraic arguments
presented here, however, reproduce the correct answer.

\section{Linear State-Based Input-Output Systems}

We obtain a linear dynamical model in the case where our system is an
assembly of quantum modes $a_{\alpha }$ $\left( \alpha =1,\cdots ,m\right) $
and the components of the generator take the special form $S_{ij}$ scalars,
(using summation convention for repeated indices from now on) 
\begin{eqnarray}
L_{i} &=&C_{i\alpha }^{-}a_{\alpha }+C_{i\alpha }^{+}a_{\alpha }^{\ast }, \\
H &=&\omega _{\alpha \beta }^{-}a_{\alpha }^{\ast }a_{\beta }+\frac{1}{2}%
\omega _{\alpha \beta }^{+}a_{\alpha }^{\ast }a_{\beta }^{\ast }+\frac{1}{2}%
\omega _{\alpha \beta }^{+\ast }a_{\alpha }a_{\beta }.
\end{eqnarray}
In this case it is possible to apply transform techniques to the dynamical
equations.\ We define the transform fields 
\begin{equation}
b\left[ s\right] \triangleq \int_{0}^{\infty }e^{-st}b\left( t\right) dt.
\label{TB}
\end{equation}
Note that 
\begin{equation}
b^{\ast }\left[ s\right] =(\int_{0}^{\infty }e^{-s^{\ast }t}b\left( t\right)
dt)^{\ast }=b\left[ s^{\ast }\right] ^{\ast }.
\end{equation}
Setting $b_{\text{in},i}\left[ s\right] =\int_{0}^{\infty }e^{-st}b_{\text{in%
},i}\left( t\right) dt$, etc., we then obtain an input-output relation of
the form 
\begin{equation}
b_{\text{out},i}\left[ s\right] =\Xi _{ij}^{-}\left( s\right) b_{\text{in}}%
\left[ s\right] +\Xi _{ij}^{+}\left( s\right) b_{\text{in},j}^{\ast }\left[ s%
\right] ,  \label{IOTF}
\end{equation}
where $\Xi _{ij}^{\mp }\left( s\right) $ are the transfer functions. (Here
we ignore additional terms involving the system modes at initial time. This
omission is justified when the model is stable.)

It is convenient to introduce the doubled up notation: for a vector $%
x=\left( x_{1},\cdots ,x_{N}\right) ^{\top }$, we write $\breve{x}=\left(
x_{1},\cdots x_{N},x_{1}^{\ast },\cdots ,x_{N}^{\ast }\right) ^{\top }$
where $\top $ is transposition, for $N\times M$ matrices $A,B$ we write $%
\Delta \left( A,B\right) =\left[ 
\begin{array}{cc}
A & B \\ 
B^{\sharp } & A^{\sharp }
\end{array}
\right] $ where $\sharp $ is entry-wise conjugation, $\left[ A_{ij}\right]
^{\sharp }=\left[ A_{ij}^{\ast }\right] $. We also set $\Delta \left(
A,B\right) ^{\flat }=\Delta \left( A^{\dag },-B^{\top }\right) $ where \dag\
is the usual hermitian conjugation. We say that a matrix $\tilde{S}=\Delta
\left( S_{-},S_{+}\right) $ is \textit{Bogoliubov}, or symplectic, if it is
invertible with 
\begin{equation*}
\tilde{S}^{\flat }=\tilde{S}^{-1}.
\end{equation*}

The transfer relation can be written as 
\begin{equation*}
\breve{b}_{\text{out}}\left[ s\right] =\tilde{\Xi}\left( s\right) \breve{b}_{%
\text{in}}\left[ s\right]
\end{equation*}
with transfer matrix function 
\begin{equation*}
\tilde{\Xi}\left( s\right) =\left[ 
\begin{array}{cc}
\Xi ^{-}\left( s\right) & \Xi ^{+}\left( s\right) \\ 
\Xi ^{+}\left( s\right) ^{\sharp } & \Xi ^{-}\left( s\right) ^{\sharp }
\end{array}
\right] .
\end{equation*}
Explicitly 

\begin{equation}
\tilde{\Xi}\left( s\right) =\left[ I_{2n}-\tilde{C}\left( sI_{2m}-\tilde{A}%
\right) ^{-1}\tilde{C}^{\flat }\right] \tilde{S}  \label{TFform}
\end{equation}
where $\tilde{S}=\Delta \left( S,0\right) $, $\tilde{C}=\Delta \left(
C_{-},C_{+}\right) $ where $C_{\mp }=\left[ C_{i\alpha }^{\mp }\right] $, $%
\tilde{A}=-\frac{1}{2}\tilde{C}^{\flat }\tilde{C}-i\tilde{\Omega}$ where $%
\tilde{\Omega}=\Delta \left( \Omega _{-},\Omega _{+}\right) $ with $\Omega
_{\mp }=\left[ \omega _{\alpha \beta }^{\mp }\right] $.

\subsection{Analysis of the Spectrum}

In addition to the fields $b\left[ s\right] $ in $\left( \ref{TB}\right) $
we also define past-field transforms 
\begin{equation}
c\left[ s\right] \triangleq \int_{-\infty }^{0}e^{-st}b\left( t\right) dt
\end{equation}
We remark that for input process as arguments, the fields $b\left[ s\right] $
and $b^{\ast }\left[ s\right] $ commute with the fields $c\left[ s^{\prime }%
\right] $ and $c^{\ast }\left[ s^{\prime }\right] $ for all parameters $%
s,s^{\prime }$ since they involve integrals over future and past input
fields respectively.

The Fourier transform of a field $b$ is then defined to be 
\begin{eqnarray*}
\hat{b}\left( \omega \right) &=&\frac{1}{\sqrt{2\pi }}\int_{-\infty
}^{\infty }e^{i\omega t}b\left( t\right) dt \\
&\equiv &\frac{1}{\sqrt{2\pi }}b\left[ 0^{+}-i\omega \right] +\frac{1}{\sqrt{%
2\pi }}c\left[ 0^{-}-i\omega \right] .
\end{eqnarray*}

The canonical commutation relations (\ref{CCR}) then imply that $\left[ \hat{%
b}_{\text{in,}i}\left( \omega \right) ,\hat{b}_{\text{in,}j}\left( \omega
^{\prime }\right) ^{\ast }\right] =\delta _{ij}\delta \left( \omega -\omega
^{\prime }\right) $. In the vacuum state we have 
\begin{eqnarray*}
\langle b_{\text{in,}i}\left[ 0^{+}-i\omega \right] b_{\text{in,}j}^{\ast }%
\left[ 0^{+}-i\omega ^{\prime }\right] \rangle &=&\delta _{ij}\zeta
_{+}\left( \omega +\omega ^{\prime }\right) , \\
\langle c_{\text{in,}i}\left[ 0^{-}-i\omega \right] c_{\text{in,}j}^{\ast }%
\left[ 0^{-}-i\omega ^{\prime }\right] \rangle &=&\delta _{ij}\zeta
_{-}\left( \omega +\omega ^{\prime }\right) ,
\end{eqnarray*}
where the Heitler functions are 
\begin{equation*}
\zeta _{+}\left( \omega \right) =\int_{0}^{\infty }e^{i\omega t}dt,\,\,\zeta
_{-}\left( \omega \right) =\int_{-\infty }^{0}e^{i\omega t}dt,
\end{equation*}
or 
\begin{equation*}
\zeta _{\pm }\left( \omega \right) =\pi \delta \left( \omega \right) \pm iPV%
\frac{1}{\omega }.
\end{equation*}
In practice we shall only encounter the combination $\zeta _{+}+\zeta
_{-}=2\pi \delta $ when calculating physical correlations, and not encounter
the principle value contribution. In particular, 
\begin{equation}
\langle \hat{b}_{\text{in,}i}\left( \omega \right) \hat{b}_{\text{in,}%
j}^{\ast }\left( \omega ^{\prime }\right) \rangle =\delta _{ij}\delta \left(
\omega -\omega ^{\prime }\right) ,  \label{bfin}
\end{equation}
as we have the sum of $\frac{1}{2\pi }\langle b_{\text{in,}i}\left[
0^{+}-i\omega \right] b_{\text{in,}j}^{\ast }\left[ 0^{+}+i\omega ^{\prime }%
\right] \rangle $ and $\frac{1}{2\pi }\langle c_{\text{in,}i}\left[
0^{-}-i\omega \right] c_{\text{in,}j}^{\ast }\left[ 0^{-}+i\omega ^{\prime }%
\right] \rangle $. Likewise 
\begin{eqnarray*}
\langle \hat{b}_{\text{in,}i}^{\ast }\left( \omega \right) \hat{b}_{\text{in,%
}j}\left( \omega ^{\prime }\right) \rangle =\langle \hat{b}_{\text{in,}%
i}\left( \omega \right) \hat{b}_{\text{in,}j}\left( \omega ^{\prime }\right)
\rangle \\
=\langle \hat{b}_{\text{in,}i}^{\ast }\left( \omega \right) \hat{b}_{%
\text{in,}j}^{\ast }\left( \omega ^{\prime }\right) \rangle =0.
\end{eqnarray*}

Ignoring the contribution from the initial value of the internal mode, the
input-output relations for the past fields takes a similar form to (\ref
{IOTF}) namely 
\begin{equation*}
c_{\text{out},i}\left[ s\right] =\Xi _{ij}^{-}\left( s\right) c_{\text{in},j}%
\left[ s\right] +\Xi _{ij}^{+}\left( s\right) c_{\text{in},j}^{\ast }[s],
\end{equation*}
the only essential difference in the calculation being the sign change. Let
us introduce the matrices 
\begin{equation}
\mathcal{S}_{ij}^{-}\left( \omega \right) =\Xi _{ij}^{-}\left( -i\omega
\right) ,\quad \mathcal{S}_{ij}^{+}\left( \omega \right) =\Xi
_{ij}^{+}\left( -i\omega \right)
\end{equation}
then 
\begin{equation}
\hat{b}_{\text{out},i}\left( \omega \right) =\mathcal{S}_{ij}^{-}\left(
\omega \right) \hat{b}_{\text{in},j}\left( \omega \right) +\mathcal{S}%
_{ij}^{+}\left( \omega \right) \hat{b}_{\text{in},j}\left( -\omega \right)
^{\ast }.  \label{bfio}
\end{equation}

We may therefore determine the correlation functions from the transfer
functions given that the input is in the vacuum state: 
\begin{eqnarray}
\left\langle \hat{b}_{\text{out},i}^{\ast }\left( \omega \right) \hat{b}_{%
\text{out},j}\left( \omega ^{\prime }\right) \right\rangle &=&\mathcal{N}%
_{ij}\left( \omega \right) \,\delta \left( \omega -\omega ^{\prime }\right) ,
\notag \\
\left\langle \hat{b}_{\text{out},i}\left( \omega \right) \hat{b}_{\text{out}%
,j}\left( \omega ^{\prime }\right) \right\rangle &=&\mathcal{M}_{ij}\left(
\omega \right) \,\delta \left( \omega +\omega ^{\prime }\right) ,
\end{eqnarray}
where 
\begin{equation}
\mathcal{N}_{ij}\left( \omega \right) =\mathcal{S}_{ik}^{+}\left( \omega
\right) ^{\ast }\mathcal{S}_{jk}^{+}\left( \omega \right) ,\;\mathcal{M}%
_{ij}\left( \omega \right) =\mathcal{S}_{ik}^{-}\left( \omega \right) 
\mathcal{S}_{jk}^{+}\left( -\omega \right) .
\end{equation}
We note the straightforward identities
\begin{equation*}
\mathcal{N}_{ij}\left( \omega \right) ^{\ast }=\mathcal{N}_{ji}\left( \omega
\right) ,\quad \mathcal{M}_{ij}\left( \omega \right) ^{\ast }=\mathcal{M}%
_{ji}\left( -\omega \right) .
\end{equation*}

We remark that 
\begin{equation*}
\tilde{\Xi}\left( -i\omega \right) \equiv \Delta \left( \mathcal{S}%
_{-}\left( \omega \right) ,\mathcal{S}_{+}\left( \omega \right) \right)
\end{equation*}
and that this defines a Bogoliubov matrix for each real $\omega $ where it
is well-defined, see \cite{GJN09} subsection V.C. In particular, this
ensures that the transformation from inputs to outputs is canonical, and the
Fourier transform of the outputs satisfy a similar relation to $\left( \ref
{bfin}\right) $. We see directly from $\left( \ref{TFform}\right) $ that $%
\lim_{|\omega |\rightarrow \infty }\tilde{\Xi}\left( -i\omega \right)
=\Delta (S,0)$, or 
\begin{equation}
\lim_{|\omega |\rightarrow \infty }\mathcal{S}_{-}\left( \omega \right)
=S,\;\lim_{|\omega |\rightarrow \infty }\mathcal{S}_{+}\left( \omega \right)
=0.  \label{S asymptotic}
\end{equation}

\bigskip

\textbf{Definition:} \textit{We say that a component is capable of spectral
squeezing if the matrix} $\mathcal{M} \left( \omega \right) $ \textit{is non zero
for certain frequencies }$\omega $. \textit{In particular, given a vacuum
input, we say that the }$i$\textit{th mode is spectrally squeezed if} $%
\mathcal{M}_{ii}\left( \omega \right) \neq 0$ \textit{for some} $\omega $.

\bigskip

In the single input situation, the $\mathcal{S}_{\mp }\left( \omega \right) $
are complex-valued functions satisfying $|\mathcal{S}_{-}\left( \omega
\right) |^{2}-|\mathcal{S}_{+}\left( \omega \right) |^{2}=1$. We then define
the \textit{spectral squeezing function} $r\left( \omega \right) $ by $|%
\mathcal{S}_{-}\left( \omega \right) |=\cosh r\left( \omega \right) $, that
is 
\begin{eqnarray}
r\left( \omega \right) &=& \frac{1}{2}\ln \frac{|\mathcal{S}_{-}\left( \omega
\right) |+|\mathcal{S}_{+}\left( \omega \right) |}{|\mathcal{S}_{-}\left(
\omega \right) |-|\mathcal{S}_{+}\left( \omega \right) |} \notag \\
&\equiv & \ln \left\{ |%
\mathcal{S}_{-}\left( \omega \right) |+|\mathcal{S}_{+}\left( \omega \right)
|\right\} .  \label{r_omega}
\end{eqnarray}

\subsection{Power Spectrum Density}

We define output quadratures by 
\begin{equation}
q_{\text{out},i}\left( t,\theta \right) =e^{i\theta }b_{\text{out},i}\left(
t\right) +e^{-i\theta }b_{\text{out},i}\left( t\right) ^{\ast },
\end{equation}
for fixed phases $\theta \in \lbrack 0,2\pi )$. The integrated processes $Q_{%
\text{out},i}\left( t,\theta \right) =\int_{0}^{t}q_{\text{out},i}\left(
t^{\prime },\theta \right) dt^{\prime }$ are self-commuting for fixed $%
\theta $ and different times $t$ and indices $i$, and correspond to classical
diffusion processes with It\={o} differentials satisfying 
\begin{equation}
dQ_{\text{out},i}\left( t,\theta \right) \,dQ_{\text{out},j}\left( t,\theta
\right) =\delta _{ij}dt.  \label{dQ2}
\end{equation}
Following Barchielli and Gregoratti \cite{BG07}, we set 
\begin{eqnarray*}
\mathcal{P}_{ij}(\omega ,\theta ,T) &=&\frac{1}{T}\langle
\int_{0}^{T}e^{i\omega t_{1}}q_{\text{out},i}\left( t_{1},\theta \right)
dt_{1} \\
&&\times \int_{0}^{T}e^{-i\omega t_{2}}q_{\text{out},i}\left( t_{2},\theta
\right) dt_{2}\rangle , \\
\mathcal{P}_{ij}^{\text{el}}(\omega ,\theta ,T) &=&\frac{1}{T}\langle
\int_{0}^{T}e^{i\omega t_{1}}q_{\text{out},i}\left( t_{1},\theta \right)
dt_{1}\rangle  \\
&&\times \langle \int_{0}^{T}e^{-i\omega t_{2}}q_{\text{out},i}\left(
t_{2},\theta \right) dt_{2}\rangle , \\
\mathcal{P}_{ij}^{\text{inel}}(\omega ,\theta ,T) &=&\mathcal{P}_{ij}(\omega
,\theta ,T)-\mathcal{P}_{ij}^{\text{el}}(\omega ,\theta ,T),
\end{eqnarray*}
and define the \emph{power spectral density} matrix to be
\begin{equation}
\mathcal{P}_{ij}(\omega ,\theta )=\lim_{T\rightarrow \infty }\mathcal{P}%
_{ij}(\omega ,\theta ,T),
\end{equation}
whenever the limits exist, along with the elastic and inelastic components $%
\mathcal{P}_{ij}^{\text{el}}(\omega ,\theta )=\lim_{T\rightarrow \infty }%
\mathcal{P}_{ij}^{\text{el}}(\omega ,\theta ,T)$, $\mathcal{P}_{ij}^{\text{%
inel}}(\omega ,\theta )=\lim_{T\rightarrow \infty }\mathcal{P}_{ij}^{\text{%
inel}}(\omega ,\theta ,T)$ respectively.

The It\={o} rule $\left( \ref{dQ2}\right) $ implies that
\begin{equation}
\mathcal{P}_{ij}^{\text{inel}}(\omega ,\theta )=\delta _{ij},
\end{equation}
and this may be interpreted by saying that the squeezing in the dynamic model comes entirely from the elastic component, and that there is no inelastic squeezing.

The Fourier transform is then $\hat{q}_{\text{out},i}\left( \omega ,\theta
\right) =e^{i\theta }\hat{b}_{\text{out},i}\left( \omega \right)
+e^{-i\theta }\hat{b}_{\text{out},i}\left( -\omega \right) ^{\ast }$ and it
is readily verified that, for vacuum input, 
\begin{equation}
\langle \hat{q}_{\text{out},i}\left( \omega ,\theta \right) \hat{q}_{\text{%
out},j}\left( \omega ^{\prime },\theta \right) \rangle =\mathcal{P}%
_{ij}\left( \omega ,\theta \right) \delta \left( \omega +\omega ^{\prime
}\right) 
\end{equation}
where we obtain the explicit expression 
\begin{eqnarray}
\mathcal{P}_{ij}\left( \omega ,\theta \right)  &=&\delta _{ij}+\mathcal{N}%
_{ij}\left( -\omega \right) +\mathcal{N}_{ji}\left( \omega \right)   \notag
\\
&+&e^{2i\theta }\mathcal{M}_{ji}\left( \omega \right) +e^{-2i\theta }\mathcal{M}%
_{ij}\left( -\omega \right) ^{\ast }.
\end{eqnarray}

\subsection{Idealized Static Squeezing Components}

A static squeezing device is an idealized static component with input-output
relation of the form (either in the time or transform domain) 
\begin{equation}
b_{\text{out},i}=S_{ij}^{-}b_{\text{in},j}+S_{ij}^{+}b_{\text{in},j}^{\ast }
\label{H}
\end{equation}
where $S_{\mp }=\left[ S_{ij}^{\mp }\right] \in \mathbb{C}^{n\times n}$ are
constant coefficients such that $\tilde{S}=\Delta \left( S_{-},S_{+}\right) $
is a Bogoliubov matrix. The outputs are then a symplectic transformation of
the inputs and therefore satisfy the canonical commutation relations.

In practice, such a device is realized approximately by a dynamical
component, in a limiting regime. Specifically, we would require in the
Fourier domain that the coefficients $\mathcal{S}_{ij}^{\mp }\left( \omega
\right) $ in $\left( \ref{bfio}\right) $ are approximately constant over a
wide range of frequencies.

\bigskip

\textbf{Definition:} \textit{We say that a sequence of models converges
pointwise in transfer function if we have} $\lim_{k\rightarrow \infty }%
\tilde{\Xi}_{k}\left( \omega \right) =\tilde{\Xi}\left( \omega \right) $.

\bigskip

If the limit is a Bogoliubov matrix $\tilde{S}=\Delta \left(
S_{-},S_{+}\right) $ independent of $s$, then we obtain a static device. In
this case, if $S_{+}=0$ then $S_{-}$ is unitary and the limit corresponds to
a beam splitter with $S=S_{-}$. The situation $S_{+}\neq 0$ can however
arise as such limits, the DPA is an example, and we refer to such idealized
components as \textit{static squeezing devices}. This notion of convergence
is weak since there is no quantum stochastic limit model for which we could
obtain $S_{+}\neq 0$, specifically we would violate the requirement $\left( 
\ref{S asymptotic}\right) $ common to all dynamical models considered up to
this point.

For the case of a single input ($n=1)$, the $S_{\mp }$ are scalars with the
constraint $|S_{-}|^{2}-|S_{+}|^{2}=1,$which ensures preservation of the
canonical commutation relations. The parameter $r=\cosh |S_{-}|$ is referred
to as the \emph{squeezing parameter}. We then have $|S_{-}|=\cosh r$, $%
|S_{+}|=\sinh r$, and we find that the extremal squeezing ratios of
quadratures by the device are $e^{\pm r}$.

We should remark that the canonical transformation in equation (\ref{H}) is
a Bogoliubov transformation for a quantum field. There is a strict condition
on when Bogoliubov transformations are unitarily implemented for infinite
dimensional systems (Shale's Theorem, \cite{Shale}) which are not met in
this particular case.

\section{The Degenerate Parametric Amplifier}

We now treat the specific example of a degenerate parametric amplifier.

\subsection{Lossy DPA, Open Loop}

We consider $n=2$ input field processes driving a single $\left( m=1\right) $
mode with 
\begin{equation*}
S=\left[ 
\begin{array}{cc}
1 & 0 \\ 
0 & 1
\end{array}
\right] ,L=\left[ 
\begin{array}{c}
\sqrt{\kappa }a \\ 
\sqrt{\gamma }a
\end{array}
\right]
\end{equation*}
and $H=H_{\text{DPA}}$ as in $\left( \ref{H_DPA}\right) $. Here $C_{-}=\left[
\begin{array}{c}
\sqrt{\kappa } \\ 
\sqrt{\gamma }
\end{array}
\right] $, $C_{+}=0$, $\Omega _{-}=0$ and $\Omega _{+}=\frac{\varepsilon }{2}$%
. Therefore $\tilde{A}=-\frac{1}{2}\left[ 
\begin{array}{cc}
\kappa +\gamma & -\varepsilon \\ 
-\varepsilon & \kappa +\gamma
\end{array}
\right] $, and we\ see that the system is Hurwitz stable (that is, $\tilde{A}
$ has all eigenvalues in the negative half plane) if 
\begin{equation}
\kappa +\gamma >\varepsilon .  \label{Hurwitz stability}
\end{equation}

We obtain the following expressions for $\Xi _{\mp }\left( s\right) =\left[
\Xi _{ij}^{\mp }\left( s\right) \right] $:

\begin{align}
\Xi _{-}\left( s\right) & =  \notag \\
& \frac{1}{P\left( s\right) }\left[ 
\begin{array}{cc}
 s^2+\gamma s +\frac{\gamma^2 -\kappa^2 -\varepsilon^2}{4}  & -\sqrt{%
\kappa \gamma }\left( s+\frac{\kappa +\gamma }{2}\right)  \\ 
- \sqrt{\kappa \gamma }\left( s+\frac{\kappa +\gamma }{2}\right)  & 
s^2+\kappa s +\frac{\kappa^2  -\gamma^2 -\varepsilon^2}{4}  
\end{array}
\right] ,  \notag \\
\Xi _{+}\left( s\right) & =-\frac{\varepsilon }{2P\left( s\right) }\left[ 
\begin{array}{cc}
\kappa  & \sqrt{\kappa \gamma } \\ 
\sqrt{\kappa \gamma } & \gamma 
\end{array}
\right] ,  \label{TF}
\end{align}

where 
\begin{equation}
P\left( s\right) =\left( s+\frac{\kappa +\gamma }{2}+\frac{\varepsilon }{2}%
\right) \left( s+\frac{\kappa +\gamma }{2}-\frac{\varepsilon }{2}\right) .
\label{P}
\end{equation}

This gives the transfer function for the component in FIG.3 prior to
feedback connection.

\subsection{Lossy DPA, Closed Loop}

Let us take for definiteness the beam splitter matrix to be 
\begin{equation}
T\left( \alpha \right) =\left[ 
\begin{array}{cc}
\alpha & \beta \\ 
\beta & -\alpha
\end{array}
\right] ,
\end{equation}
where $0<\alpha <1$ and $\beta =\sqrt{1-\alpha ^{2}}$. Following our
discussions in the previous section, the actual situation modeled in FIG.3
is then given by $_{{}}$replacing $L_{1}$ by 
\begin{equation}
L_{1}\left( \alpha \right) =\beta \left( 1-\nu \right) L_{1}\equiv \sqrt{%
\kappa \left( \alpha \right) }a
\end{equation}
where $\kappa \left( \alpha \right) =\frac{1-\alpha }{1+\alpha }\kappa $ in
accordance with $\left( \ref{kappa_eff}\right) $. The transfer function is
therefore of the same form as derived in $\left( \ref{TF},\ref{P}\right) $
but with $\kappa $ now replaced by $\kappa \left( \alpha \right) $.

\subsection{Spectrum of the DPA Output}

We are interested in the input-output relation between $b_{\text{in,1}}$ and 
$b_{\text{out,1}}$. Here, displaying the dependence on $\alpha $, we compute 
\begin{widetext}
\begin{equation}
\mathcal{N}_{11}\left( \omega ,\alpha \right) =\frac{\varepsilon ^{2}\kappa (\alpha ) \left(
\kappa (\alpha  ) +\gamma \right) }{4 D(\omega ,\alpha  )}, \quad
\mathcal{M}_{11}\left( \omega ,\alpha  \right) = \frac{\varepsilon \kappa (\alpha  ) }{%
2 D(\omega ,\alpha )}
\left[ \omega^{2}+\left( \frac{\kappa (\alpha )+\gamma }{2}\right) ^{2}+\left( \frac{\varepsilon }{%
2}\right) ^{2}\right] 
\label{N_M}
\end{equation}
with 
\begin{equation}
D(\omega ,\alpha  ) =|P\left( -i\omega ,\alpha  \right) |^{2}=\left[ \omega ^{2}+\left( \frac{\kappa (\alpha )
+\gamma +\varepsilon }{2}\right) ^{2}\right] \left[ \omega ^{2}+\left( \frac{%
\kappa (\alpha  ) +\gamma -\varepsilon }{2}\right) ^{2}\right] .
\label{D}
\end{equation}
\end{widetext}

Note that in the lossless situation $\gamma =0$, we have $|\mathcal{S}%
_{11}^{-}\left( \omega ,\alpha \right) |^{2}-|\mathcal{S}_{11}^{+}\left( \omega ,\alpha 
\right) |^{2}=1$ and therefore the identity $|\mathcal{M}_{11}(\omega ,\alpha 
)|^{2}=(\mathcal{N}_{11}(\omega ,\alpha  )+1)\mathcal{N}_{11}(\omega ,\alpha )$. In
particular, we compute that the spectral squeezing function is 
\begin{equation}
r\left( \omega ,\alpha \right) =\frac{1}{2}\ln \frac{\omega ^{2}+\left( \frac{\kappa (\alpha  )
+\varepsilon }{2}\right) ^{2}}{\omega ^{2}+\left( \frac{\kappa (\alpha  )-\varepsilon 
}{2}\right) ^{2}}.
\end{equation}
In this case we find that we find that the power spectral density $\mathcal{P}_{11}\left( \omega ,\theta \right)$ is 
\begin{equation*}
 \frac{ \left[ \omega ^{2}+\frac{\kappa (\alpha ) ^{2}+\varepsilon ^{2}}{4}%
\right] ^{2}+\frac{\kappa (\alpha) ^{2}\varepsilon ^{2}}{4}+\varepsilon \kappa (\alpha )\left[
\omega ^{2}+\frac{\kappa (\alpha )^{2}+\varepsilon ^{2}}{4}\right] \cos 2\theta
}{D\left( \omega ,\alpha
\right) } ,
\end{equation*}
with the maximum squeezing at $\theta =0$, $\mathcal{P}_{11}\left( \omega
,0\right) =e^{2r (\alpha , \omega ) }$ and minimum squeezing at $\theta
=\frac{\pi}{2} $, $\mathcal{P}_{11}\left( \omega ,\frac{\pi}{2}\right) =e^{-2r (\alpha , \omega
) }$.

\subsection{The Static Limit of the DPA}

We consider a sequence of DPA models described by the parameters $\left(
\kappa _{k},\varepsilon _{k},\gamma _{k}\right) _{k\geq 1}$ with 
\begin{equation*}
\kappa _{k}=k\kappa ,\;\varepsilon _{k}=k\varepsilon ,\;\gamma _{k}=k\gamma ,
\end{equation*}
and consider the singular limit $k\rightarrow \infty $. We note that $\Xi
_{\mp }^{\left( k\right) }\left( s\right) \equiv \Xi _{\mp }\left(
s/k\right) $ so the limit is equivalent to the low frequency limit. The
limit transfer functions are independent of the transform variable $s$: 
\begin{eqnarray*}
S_{-} &=&\lim_{k\rightarrow \infty }\Xi _{-}^{\left( k\right) }\left(
s\right)  \\
&=&\frac{1}{\left( \kappa +\gamma \right) ^{2}-\varepsilon ^{2}}\left[ 
\begin{array}{cc}
\gamma ^{2}-\kappa ^{2}-\varepsilon ^{2} & -2\sqrt{\kappa \gamma }\left(
\kappa +\gamma \right)  \\ 
-2\sqrt{\kappa \gamma }\left( \kappa +\gamma \right)  & \kappa ^{2}-\gamma
^{2}-\varepsilon ^{2}
\end{array}
\right] , \\
S_{+} &=&\lim_{k\rightarrow \infty }\Xi _{+}^{\left( k\right) }\left(
s\right) =\frac{-2\varepsilon }{\left( \kappa +\gamma \right)
^{2}-\varepsilon ^{2}}\left[ 
\begin{array}{cc}
\kappa  & \sqrt{\kappa \gamma } \\ 
\sqrt{\kappa \gamma } & \gamma 
\end{array}
\right] .
\end{eqnarray*}
In particular, $\tilde{S}=\Delta \left( S_{-},S_{+}\right) $ is a Bogoliubov
matrix. The squeezing parameter for the limit Bogoliubov transformation is
then 
\begin{equation*}
r=\ln \left( \frac{\kappa +\varepsilon }{\kappa -\varepsilon }\right) ,
\end{equation*}
or equivalently $r\left( 0\right) $ in $\left( \ref{r_omega}\right) $. This
is of course the equation $\left( \ref{r_open_loop}\right) $.

The central issue here is that the asymptotic limit $|\omega |\rightarrow
\infty $ and the transfer function convergence limit $ k \rightarrow
\infty $ do not commute:
denoting the spectral squeezing functions for the sequence of models as $%
r^{\left( k\right) }\left( \omega \right) $ we have $\lim_{k\rightarrow
\infty }r^{\left( k\right) }\left( \omega \right) =r$ for all $\omega $,
while $\lim_{|\omega |\rightarrow \infty }r^{\left( k\right) }\left( \omega
\right) =0 $ for all $k$.

We likewise find that the inelastic contribution to the power spectrum of the limit output quadratures
is given by $\left( \theta =0\right) $
\begin{equation*}
\mathcal{P}^{\text{inel}}\left( \theta =0\right) = \frac{\left( \kappa
+\gamma +\varepsilon \right) ^{2}}{\left[ \left( \kappa +\gamma \right)
^{2}-\varepsilon ^{2}\right] ^{2}} \times
\end{equation*}
\begin{equation}
\left[ 
\begin{array}{cc}
( \kappa +\gamma ) ^{2}+\varepsilon
^{2} -2\varepsilon ( \gamma -\kappa )  & 4\sqrt{\kappa \gamma }\varepsilon  \\ 
4\sqrt{\kappa \gamma} \varepsilon  & (\kappa +\gamma )^{2}+\varepsilon ^{2} +2\varepsilon( 
\gamma  - \kappa )
\end{array}
\right] 
\end{equation}
which has eigenvalues unity and $\frac{\left( \kappa +\gamma +\varepsilon
\right) ^{4}}{\left[ \left( \kappa +\gamma \right) ^{2}-\varepsilon ^{2}%
\right] ^{2}}$ and is therefore positive definite as required. The matrix
for $\theta =\frac{\pi }{2}$ is obtained by replacing $\varepsilon $ by $%
-\varepsilon $. We have seen that for the dynamic approximation we always
have $\mathcal{P}_{k}^{\text{inel}}\left( \omega ,\theta \right) =I$, the
identity matrix, for finite $k$. The limit situation on the contrary now has purely inelastic squeezing.

\section{Feedback-Enhanced Squeezing}

For an idealized static description of a lossless DPA when placed in loop as
in FIG.3, we find that the squeezing parameter is modified to 
\begin{equation*}
r_{\alpha }=\ln \left( \frac{\kappa \left( \alpha \right) +\varepsilon }{%
\kappa \left( \alpha \right) -\varepsilon }\right) ,
\end{equation*}
with $\kappa \left( \alpha \right) =\frac{1-\alpha }{1+\alpha }\kappa $. We
observe that the critical value of the reflectivity $\alpha $ is 
\begin{equation}
\alpha _{\text{crit}}=\frac{\kappa -\varepsilon }{\kappa +\varepsilon }.
\end{equation}
Here $\kappa \left( \alpha _{\text{crit}}\right) =\varepsilon $, and the
squeezing parameter diverges. The approximating dynamical model has spectral
squeezing function 
\begin{equation}
r\left( \omega ,\alpha_{\text{crit}} \right) =\frac{1}{2}\ln 
\frac{\omega ^{2}+\varepsilon^2}{\omega^2 }
.
\end{equation}
which possesses a logarithmic singularity at $\omega =0$ for the critical
situation. The open-loop system is stable if and only if $\kappa
>\varepsilon $, while for the closed-loop system this is modified to $\kappa
\left( \alpha \right) >\varepsilon $. Therefore the infinite squeezing
situation implies the onset of instability of the closed-loop amplifier.

\bigskip 

It is instructive to look at the lossy $\left( \gamma >0\right) $
closed-loop situation. The relevant description is then given by (\ref{N_M}) and (\ref{D}). Hurwitz stability requires $\kappa \left( \alpha \right) +\gamma
>\varepsilon $. We assume that the dissipation is below the threshold value $%
\left( \gamma <\varepsilon \right) $ and that the open-loop system is stable 
$\left( \kappa +\gamma >\varepsilon \right) $, then the closed loop system
is stable for $\alpha \in \left( 0,\alpha _{\text{crit}}\right) $, where the
critical value is now
\begin{equation}
\alpha _{\text{crit}}=\frac{\kappa -\varepsilon +\gamma }{\kappa
+\varepsilon -\gamma }.
\end{equation}
Here the critical value solves $\kappa \left( \alpha _{\text{crit}}\right)
=\varepsilon -\gamma $, and we have
\begin{eqnarray}
\mathcal{N}_{11}\left( \omega ,\alpha _{\text{crit}}\right)  &=&\frac{%
\varepsilon ^{3}\left[ \varepsilon -\gamma \right] }{4\left[ \omega
^{2}+\varepsilon ^{2}\right] \omega ^{2}}, \\
\mathcal{M}_{11}\left( \omega ,\alpha _{\text{crit}}\right)  &=&\frac{%
\varepsilon \left[ \varepsilon -\gamma \right] \left[ \omega^{2}+ 
\left( \frac{\varepsilon }{2}\right) ^{2} \right]}{2\left[ \omega ^{2}+\varepsilon ^{2}%
\right] \omega ^{2}}.
\end{eqnarray}
which are finite for $\omega \neq 0$. Both expressions diverge in the limit $%
k\rightarrow \infty $, however, when we replace the parameters by $\left(
\kappa _{k},\varepsilon _{k},\gamma _{k}\right) $.

\bigskip

\section{Conclusion}
Coherent quantum feedback control offers an intriguing potential to engineer physically interesting states, and achieve high performance for quantum devices. The feedback approach based on quantum measurement is limited to time scales set by the measurement apparatus, the computer estimating (filtering) the quantum state of the system based on the measurements, and the implementation of the controls by the actuator based on the filtered state. In contrast, coherent control is limited by the time delays associated in light traversing the loop.

We have shown that coherent feedback control can enhance the capability of a device to squeeze quadratures by using an optical network involving a beam-splitter loop. Conversely, squeezing could be suppressed by altering the beam splitter, for instance, by reversing the sign of $\alpha$, though there would arguably only enhancement would be desirable.

By tuning the beam-splitter reflectivity, we can modify the effective damping of an in-loop degenerate parametric amplifier while leaving the  amplifier
The situation where the squeezing becomes infinite corresponds to the threshold value of the damping as discussed by Gardiner \cite{GZ00}, however, we observe that this

\begin{acknowledgments}
We acknowledge several highly valuable comments from Professors Matthew James, Robin Hudson, Rolf Gohm, Ben Shuttleworth, Hendra Nurdin and Masahiro Yanagisawa. We are particularly grateful to the referee for several suggestions that have improved immensely the background and scope of the paper.
\end{acknowledgments}


\end{document}